\documentstyle[prd,aps]{revtex}
\begin{document}

\preprint{
   BI-TP 96/1}
%\draft
%
%%%%%%%%%%%%%%%%%%%%%%%%
\title{Anomalous Processes in Effective Chiral Theories\\
at High Temperatures}

\author{R.~Baier, M.~Dirks\footnote{Supported by 
Deutsche Forschungsgemeinschaft}
, and O.~Kober}
\address{
Fakult\"{a}t f\"{u}r Physik, Universit\"{a}t Bielefeld,
D-33501 Bielefeld, Germany
}
%\date{\today}
\date{March 5, 1996}

\maketitle

%%%%%%%%%%%%%%%%%%%%%%%%%%%%%%%%%

\begin{abstract}
We discuss chiral theories of constituent quarks interacting
with bosons at high temperatures.
In the chirally symmetric phase we demonstrate
by applying functional
methods the presence of effective anomalous couplings for e.g.
$\pi \sigma \to \gamma \gamma, ~ 
\gamma \to \pi \pi \pi \sigma, ~K K \to \pi \pi \pi \sigma$, etc., 
as they have recently
been discussed by Pisarski.
\end{abstract}

% body of paper here
%%%%%%%%%%%%%%%%%%%%%%%%%%%%%%%%%%%%%%%%%%
%%%%%%%%%
%\newpage
%%%%%%%%%%%%%%%%%%%%%%%%%%%%%%%%%%%%%%%%%%%%
%\narrowtext
\section{Introduction}

In recent papers\cite{pisa1,pisa2} Pisarski pointed out that although
the chiral (abelian as well as nonabelian) anomaly\cite{ra,rb}
in terms of fundamental fields is temperature independent\cite{rc,rd}
the manifestation of the anomaly, however, in terms of effective fields 
changes with temperature.
When considering e.g. $\pi^0 \to 2 \gamma$ the observation in 
\cite{pisa1,pisa2}
is: "In a hot, chirally symmmetric phase, $\pi^0$ doesn't go into $2 \gamma$,
but $\pi^0 \sigma$ does"! This statement indeed contradicts results
which have been obtained previously\cite{rh,ri}.
 
In \cite{pisa1,pisa2} the effective anomalous couplings for
$\pi^0 \to 2 \gamma$ ($\pi^0 \sigma \to 2 \gamma$) are
found in the framework of the constituent quark model\cite{rb,rf}
by calculating the contribution from the Feynman one-loop
triangle (box) diagrams at temperature $T = 0$ and at nonzero,
high $T$, respectively.

In this note we attempt a different approach in order to confirm
Pisarski's interesting result.
As in \cite{pisa1,pisa2} we start from the linear sigma model with
constituent quarks interacting with bosons\cite{donoh,poko}.
In order to include the effect of the axial anomaly on the 
bosons we transform the basis of right- and left-handed
quarks following ref.\cite{manm}, but in  the situation
of the chirally symmetric phase at high $T$\cite{rj}.
 At $T = 0$ and in
the spontaneously broken phase this procedure allows a direct
calculation of the Wess-Zumino-Witten action\cite{rg}
for the Goldstone bosons, after applying functional methods for
the evaluation of the fermion determinant in
connection with path integrals\cite{fuji}.

In Sec. \ref{EChL} we define the effective lagrangian and describe 
the way of performing the chirally symmetric limit 
at high temperature. 
Sec. \ref{zetareg} is devoted to the zeta-regularization used
to evaluate the fermion determinant in the symmetric case. 
In Secs. \ref{em} and \ref{had} we discuss the  effective anomalous 
low-energy  electromagnetic and hadronic couplings   
as a result of the chiral anomaly at high $T$.

%%%%%%%%%%%%%%%%%%%
\section{Effective Chiral Lagrangians} \label{EChL}

We consider the $SU(2)_L \bigotimes SU(2)_R$ lagrangian for $N_C$
coloured right- and left-handed quarks parametrized in the linear 
form
("$ \Sigma$- basis"), 
\begin{equation}\label{L} 
{\cal L} = {\overline {\psi}}_L i {\not \! \partial} \psi_L
 + \overline{\psi}_R i {\not \! \partial} \psi_R
 -g (\overline{\psi}_L \Sigma \psi_R + 
      \overline{\psi}_R \Sigma^{\dagger} \psi_L )
 + {\cal L}_{boson}  ,
\end{equation}
with the $SU(2)$ matrix
$\Sigma = \sigma + i {\vec{\tau}} {\cdot} \vec{\pi}$,
where we closely follow the notation used in \cite{donoh}.
In the following the explicit form of the boson  ~lagrangian
${\cal L}_{boson}$ is not needed\cite{donoh,poko}.
Although in the strictly symmetric phase the constituent quark
mass $m$ has to vanish, we nevertheless introduce explicitly
a breaking term  
\begin{equation}\label{Lm}
{\cal L}_m = - m \overline{\psi} \psi  .
\end{equation}
We treat $m$ as an explicit regularization parameter, which is finally
removed in
the symmetric phase by performing the limit $m \to 0$.
However, we do not break the chiral symmetry spontaneously by the 
standard redefinition of the scalar field $\sigma$. 

Following \cite{manm} we change the quark fields by 
\begin{equation}\label{transf} 
     {\psi_L}^{\prime} \equiv \xi^{\dagger} \psi_L ~~,
 ~~~~   {\psi_R}^{\prime} \equiv \xi \psi_R ,
\end{equation}
("$\xi$- basis")
defining a unitary matrix $\xi$ in such a way that the quark-pion coupling
in Eq.~(\ref{L}) becomes replaced by a derivative coupling
(as a consequence the physical predictions change in the low-energy limit,
e.g. for $\pi^0 \to 2 \gamma$, implying "anomalous inequivalence"
as discussed  for the case $T =0$ in \cite{manm}).
 
In the symmetric phase we define for nonvanishing $m$ 
\begin{equation}\label{xi} 
\xi^2 \equiv {{1 + g/m \Sigma}
\over {(1 + g/m \Sigma)^{1 \over 2}~ 
(1 + g/m \Sigma^{\dagger})^{1\over 2}}}~. 
\end{equation}
Expanding in terms of the boson fields ($\sigma, \vec{\pi}$)
the matrix $\xi$ is equivalent (up to quadratic terms)
to 
\begin{equation}\label{xiexp} 
\xi \simeq
\exp [{ {i \vec{\tau} \cdot \vec{\pi}}
\over {2 (m/g + \sigma)}}]~. 
\end{equation}
This matrix replaces the transformation matrix in the broken phase
given by
\begin{equation}\label{xiexpT0} 
\xi = 
\exp [{ {i \vec{\tau} \cdot \vec{\pi}}
\over {2 F_{\pi}} }]~, 
\end{equation}
where $F_{\pi}$ is the pion decay constant.
In the chirally broken phase the transformation (\ref{xiexpT0}) may be
obtained from (\ref{xiexp}) by first shifting 
$\sigma \to <\sigma >+~ \sigma~ \hat{=} ~F_{\pi} + \sigma$
and then performing the well defined limit $m \to 0$, while 
the dynamical $\sigma$ field becomes heavy and is therefore 
assumed to decouple.
However, approaching the symmetric phase, where $<\sigma >$ and $F_{\pi}$
vanish, the $\sigma$ field does not decouple and the mass parameter has to
be kept as $m\neq 0$.

The transformation Eq.~(\ref{transf}) yields 
\begin{equation}\label{Lxi}
{\cal L} = {\overline {\psi}}^{\prime} (i {\not \!\! D} - m) \psi^{\prime}
 + {\cal L}_{boson}  
 + O (g \sigma \overline{\psi}^{\prime}  \psi^{\prime})~, 
\end{equation}
with 
\begin{equation}\label{D}
 D_\mu = \partial_\mu + i {\overline V}_\mu + i {\overline A}_\mu \gamma_5 ~,
\end{equation} 
where the currents \cite{donoh} are defined by 
\begin{equation}\label{VA} 
{\overline V}_\mu = {- i \over 2} (\xi^\dagger \partial_\mu \xi 
                   + \xi \partial_\mu \xi^\dagger)~~, ~~
 {\overline A}_\mu = {- i \over 2} (\xi^\dagger \partial_\mu \xi 
                   - \xi \partial_\mu \xi^\dagger)~~.
\end{equation}
Explicitly, for infinitesimal transformations $\xi$
the axial current ${\overline A}_\mu$ 
when expressed in terms  of the boson fields
becomes 
\begin{equation}\label{A}
 {\overline A}_\mu = \frac{1}{2} \frac{1}{(m/g +\sigma)^2} 
[ ({m \over g} + \sigma) \vec{\tau} \cdot \partial_\mu \vec{\pi}
 - \vec{\tau} \cdot \vec{\pi} \partial_\mu \sigma ] ~~.
\end{equation}
Keeping fixed $m\neq 0$ and expanding in the fields gives 
\begin{equation}\label{Aexpand} 
 {\overline A}_\mu \simeq \frac{1}{2} \left( \frac{g}{m} \right)^2 
[ ({m \over g} - \sigma) \vec{\tau} \cdot \partial_\mu \vec{\pi}
 - \vec{\tau} \cdot \vec{\pi} \partial_\mu \sigma ] + \cdots ~~,
\end{equation}
where the dots indicate higher than quadratic terms in the
($\sigma, \vec{\pi}$) fields. It is important
to notice that because of the factor $({g / m})^2$ the current 
${\overline A}_\mu$  becomes singular for $m \to 0$, and as a consequence
the  symmetric limit is not immediately obtained.

Although the lagrangian remains invariant under the
transformation, Eq.~(\ref{transf}), the jacobian is not unity,
 as it is well known. Thus, in order to obtain equivalent low-energy physics
in both representations, i.e. in the $"\Sigma$-" as well as in the
$"\xi$- basis", we  have to calculate  
the fermion determinant for the transformation (\ref{transf}) - at $T = 0$
leading to the Wess-Zumino-Witten functional\cite{rg} -
at high temperature, and finally we have to perform the chirally symmetric
limit for $m \to 0$.

%%%%%%%%%%%%%%%%%%%%%%%%%%%%%%%%%%
\section{Zeta- Regularization at High Temperature} \label{zetareg}

The jacobian for the transformation (\ref{transf}) of the quark fields is 
determined by standard steps. Here we closely
follow the discussion in Chapters III, VII and in Appendix B of 
\cite{donoh}.
First one defines infinitesimal transformations by introducing
a continous parameter $t$, $0 \le t \le 1$, and  extending the transformation
of Eq.~(\ref{xiexp}), 
\begin{equation}\label{xit} 
\xi \to \xi_{t} \equiv
\exp [{ {i t \vec{\tau} \cdot \vec{\pi}}
\over {2 (m/g + \sigma)}}]~ \equiv \exp [ i t \overline{\pi}]~.
\end{equation}
The effective action $\Gamma$,  i.e. the
Wess-Zumino-Witten action\cite{rg} in the broken phase, is
defined by 
\begin{equation}\label{WZW} 
 \Gamma \equiv 2 \int_0^1 dt ~ tr (\overline{\pi} \gamma_5)
\equiv  \int d^4x ~{\cal L}_{odd} ~~,
\end{equation}
where the lagrangian ${\cal L}_{odd}$ contains all the 
interactions in the odd-intrinsic-parity sector of the 
chiral model under consideration.
In the following we only consider local terms as given by 
a Taylor expansion representation in $t$ of $\Gamma$ and ${\cal L}_{odd}$,
respectively.

To regularize the trace in Eq.~(\ref{WZW})
 we apply the zeta function method\cite{zeta}
combined with the expansion for the heat kernel $H(x,\tau)$\cite{donoh}.
We start from  
\begin{equation}\label{tr} 
 tr (\overline{\pi} (x) \gamma_5) = N_C tr^{\prime}
  \int d^4x <x|\overline{\pi} (x) \gamma_5 | x> ~
   \Rightarrow ~ N_C \int d^4x ~tr^{\prime}
~( \overline{\pi} (x) \gamma_5 ~\zeta_s (x) )~,
\end{equation}
 ($tr^{\prime}$ indicates the Dirac and flavour trace)
and introduce 
\begin{equation}\label{zeta} 
 \zeta_s (x) \equiv {1 \over {\Gamma(s)}}
  \int_0^{\infty} d\tau \tau^{s - 1} H (x, \tau) ~,
\end{equation}
with the heat kernel for massive fermions 
\begin{equation}\label{H}
 H(x, \tau) \equiv <x | \exp [ - \tau (D_\mu D^\mu + m^2)] |x>~,
\end{equation}
and the derivative $D_\mu$ given by Eq.~(\ref{D}).

At $T =0$, in the broken phase, the jacobian (\ref{tr})
requires the regularization, Eq.~(\ref{zeta}), in order to suppress the
contributions of the ultraviolet modes of 
the  operator $D_\mu$\cite{fuji}.
This is achieved by the $\zeta$-function regularization 
starting with a convergent expression  $\zeta_s (x)$
 for $s > 0$, which is then analytically continued to $s = 0$.
At high $T$, i.e. in the chirally symmetric phase,
we propose to continue in the variable $s$ in such a way
that a non-trivial limit for ${\cal L}_{odd}$ with
$m \to 0$ is finally obtained.

A few technical steps have to be performed and described:

\begin{itemize}
\item{} the Dirac operator is defined by 
\begin{equation}\label{DD}
D_\mu D^\mu + m^2 = d_\mu d^\mu + \sigma(x) + m^2~,~
d_\mu = \partial_\mu + i {\overline V}_\mu (x) +
 \sigma_{\mu \nu} {\overline A}_\nu (x) \gamma_5~,
\end{equation} 
and $\sigma (x)$ as defined in \cite{donoh}.
For the following it is crucial that only even powers
of the quark mass $m$ are present in this operator\cite{fuji};

\item{} at high $T$ it is convenient to continue first
to euclidean coordinates and momenta ($p_E \equiv (\vec{p}, p_4)$);

\item{} introducing a complete set of energy (Matsubara) and 
momentum eigenstates one obtaines the expansion  
%%%%
\begin{eqnarray}\label{Hexpand} 
H &(x, \tau) & \simeq  
 T~ \Sigma \int {{d^3p} \over {(2\pi)^3}} \exp [ - \tau (p_E^2 + m^2)]
%\nonumber \\
%
~ [ {{\tau^2} \over 2} (d \cdot d - \sigma)(d \cdot d - \sigma)
\nonumber \\
 &  + & {{4 \tau^3} \over {3!}} [p_E \cdot d~ p_E \cdot d (\sigma - d \cdot d)
 + p_E \cdot d (\sigma - d \cdot d) p_E \cdot d
 + (\sigma - d \cdot d) p_E \cdot d~ p_E \cdot d ]
\nonumber \\
 & + & {{16 \tau^4} \over {4!}} [p_E \cdot d~ p_E \cdot d~ p_E \cdot d~
             p_E \cdot d] + ......]~.
\end{eqnarray}
Here we drop i) terms $O(\tau^0,\tau^1)$, which give vanishing contributions
after taking
the Dirac trace in Eq.~(\ref{WZW}), and ii) terms,
which either do not contribute at $T = 0$, or which lead to
interactions in higher orders of the boson fields at high $T$.
$\Sigma$ sums the fermionic frequencies $p_4 = 2 \pi T ( n + 1/2)$
with integer $n$;

\item{} at $T=0$ the integration over momentum  is performed using 
the replacement 
\begin{equation}\label{pp}
     p_E^\mu p_E^\nu  \rightarrow   p_E^2 ~\delta^{\mu \nu} / 4~,
\end{equation} 
and related ones (cf. \cite{donoh}). At $T\neq 0$ one has to distinguish 
$p_4$ and $\vec p$ components in the expression for $H(x,\tau)$, 
Eq.~(\ref{Hexpand}). We have checked, however, that the final results, 
summarized in Eqs.~(\ref{zetas-1}, \ref{zetas-3}), are identically derived 
by just using this `covariant' prescription. To simplify the presentation 
we therefore choose to continue using Eq.~(\ref{pp}) in the following;

\item{} in order to evaluate the Matsubara frequency sum we first
perform the $\tau$-integration in Eq.~(\ref{zeta})
 and then introduce the following
expressions (cf. Appendix in \cite{lands})  
\begin{equation}\label{Tint} 
T~ \Sigma \int {{d^3p} \over {(2\pi)^3}} 
{ 1 \over {(p_E^2 + m^2)^{s + \beta}}}
\equiv 
{{(T^2)^{2 -s-\beta}} \over {(2 \pi)^{2s +2\beta -3}}}
 ~I (y^2; s+ \beta, { 1\over 2})~,
\end{equation}
where $ y = m/ (2 \pi T)$ and the function $I$ are dimensionless quantities.
For $s > 0$ the function $I$ has the following series expansion\cite{lands},
in the following required  for $\beta = 2,3,...$,
\begin{eqnarray}\label{Iexpand} 
I(y^2; s + \beta, {1 \over 2})  & = &
{1 \over { 4 {\pi}^{3/2}}} { 1 \over {\Gamma(s+\beta)}}
 ~\Sigma_0^{\infty} ~ {{(-)^k} \over {k!}} \Gamma(k+s+\beta -3/2)
\nonumber \\
 & & \times \zeta(2k + 2s +2\beta -3, {1 \over 2}) ~y^{2k}~,
\end{eqnarray}
where the Hurwitz zeta-function is related by
$\zeta (z, {1 \over 2}) \equiv (2^z - 1) \zeta (z)$ to the Riemann
function $\zeta (z)$
\cite{abram}. It is crucial to realize that only
even powers in $(m/T)$ are present in Eqs.~(\ref{Tint},\ref{Iexpand});

\item{} taking care of the proper dimension of $\tau$ in Eq.~(\ref{Hexpand})
 we multiply $\zeta_s(x)$ by $T^{2s}$; furthermore we observe
that the regularization prescription Eq.~(\ref{zeta}) still allows for
an additional normalization factor $N(s)$,
however, with the constraint $N(s = 0) = 1$; below we argue that
\begin{equation}\label{N}
 N (s) =
% {2 \over {\sqrt{\pi}}} 
      {{ (2 \pi)^{2s}~ \Gamma (1/2 -s) ~\zeta(1 - 2s, {1 \over 2})}
                   \over { \Gamma (3/2)~ \Gamma ( -s)}}
\end{equation}
for $s \le 0$.
\end{itemize}

 After performing these steps the function $\zeta_s (x)$ can be written
as (in Minkowski metric) 
%%%%
\begin{eqnarray}\label{zetaexpand} 
& \zeta_s & (x)  \simeq  {{ \pi N(s)} \over {(2 \pi)^{2s + 2}~ \Gamma (s)}}
\nonumber \\
  & \times & \left( ( d \cdot d - \sigma )^2 ~ \Gamma (s +2)~
     I(y^2; s +2, {1 \over 2}) \right.
\nonumber \\ 
&   & + { 1\over 3}[ 2 d \cdot d (d \cdot d - \sigma)
    + d_{\mu} (d \cdot d - \sigma) d^{\mu}] ~ \Gamma(s + 3)~
  [ y^2 I(y^2; s +3, {1 \over 2}) - I (y^2; s + 2, {1 \over 2})]
\nonumber \\
&   & + { 1 \over {18}} [ (d \cdot d)^2 + d_{\mu} d_{\nu} d^{\mu} d^{\nu}
        + d_{\mu} (d \cdot d) d^{\mu} ]~ \Gamma (s+4) 
\nonumber \\
&   & \times \left.
    [ y^4 I ( y^2; s + 4, { 1\over 2}) - 
  2 y^2 I(y^2; s +3, {1 \over 2}) + I (y^2; s + 2, {1 \over 2})] 
+ ... \right) ~.
\end{eqnarray}

In case the functions $d_\mu$ and $\sigma$ in Eq.~(\ref{DD})
are entirely expressed in terms of fundamental fields, which occurs
when studying the divergence of the axial currents in connection
with the abelian as well as nonabelian anomalies, including their
$T$-dependence, the limit $m \to 0 $ is trivial: it amounts
to take only the first term in the series of Eqs.~(\ref{Iexpand},\ref{zetaexpand}).
The $s$-dependence of $\zeta_s(x)$ becomes proportional to
$\zeta (2s + 1, {1 \over 2}) /\Gamma (s)$, which is only nonvanishing
when $s = 0$, due to the single pole of the Hurwitz
function $\zeta (z, {1 \over 2})$ at $z = 1$.
As a result \cite{donoh} we obtain 
\begin{equation}\label{zetas0} 
\zeta_s (x) ~ {\buildrel {s \to 0} \over \longrightarrow}~
 {1 \over {(4 \pi)^2}} ~\bigl( { 1\over 2} \sigma^2 
+ { 1 \over {12}} [ d_\mu, d_\nu ] [ d^\mu , d^\nu ]
+ { 1 \over 6} [ d_\mu [ d^\mu , \sigma]] ~  \bigr)~,
\end{equation}
independent of $T$.
Since the divergence of the anomalous (abelian and nonabelian)
axial current is directly determined by
 $tr^{\prime}~(\gamma_5 \zeta_{s=0} (x))$, where
$(d_{\mu}, \sigma)$ are expressed in terms of gauge fields\cite{ra,rb,donoh},
 the anomalies  do not depend on the
 temperature\cite{rc,rd}: the well known result is here  reproduced with
the regularization prescription under discussion. 

However, for the case of the effective theories, one first has to collect
the terms present in $\zeta_s (x)$ with different powers  in $m$,
i.e. in $y^2$, in order to finally perform the chirally symmetric limit.
The nonvanishing and regular terms to be found in this limit
give rise to anomalous interactions in the odd-intrinsic-parity sector
of the effective chiral theories at high temperatures.
These low-energy couplings are, however, different from the ones
obtained in the broken phase as pointed out in \cite{pisa1,pisa2}.

%%%%%%%%%%%%%%%%%%%%%%%%%%%%%%%%%%%%%%%%%%%%%%%%%%%

\section{Anomalous Electromagnetic Couplings} \label{em}

We now derive the anomalous coupling in leading order for
$\pi^0 \sigma ~\to 2 \gamma$ in the chirally symmetric phase.
We introduce the electromagnetic interaction by replacing
(cf. Eq.~(\ref{DD})):
$d_\mu \rightarrow d_\mu + i e Q A_\mu$, where $A_\mu$
is the photon field and $Q = 1/2~ (1/3 + \tau_3)$ (for $SU(2)$).
The thermal bath is constituted by the $\pi^0$ and $\sigma$ fields,
to which external photons are coupled, i.e. we assume that the photons are 
\underline{not} thermalized due to their small electromagnetic coupling.
In this respect we differ from \cite{pisa2}, where the photons are kept in thermal
equilibrium with $\pi^0$ and $\sigma$.

From the discussion in Sec. \ref{zetareg}  we look for 
nonvanishing contributions to  
\begin{equation}\label{L2ga1}
 {\cal L}_{T}^{2 \gamma} =
     {\lim_{m \to 0}  } ~ 
2 N_C {\int_0^1}~dt~tr^\prime~ \bigl( {{\tau_3 \pi^0} \over {2 (m/g + \sigma)}}
~\gamma_5 \zeta_s (x) \vert_{2\gamma} \bigr)~,
\end{equation} 
where the two-photon part $\zeta_s (x) \vert_{2\gamma}$ is 
provided by terms $O(\sigma^2)$ in Eq.~(\ref{zetaexpand})
as $\sigma \sim \sigma_{\mu\nu} F^{\mu\nu}$ \cite{donoh}.
Expanding in the fields $(\sigma,\vec\pi)$, e.g.
$1/(m/g+\sigma) \simeq g/m-g^2/m^2\sigma$, it is seen, that
the symmetric limit $m\to 0$ is no longer trivial in Eq.~(\ref{L2ga1}).
However, regrouping terms in Eqs.~(\ref{Iexpand},\ref{zetaexpand})
one finds an expansion with respect to $y^2=m^2/(2\pi T)^2$
with coefficients proportional to 
\begin{equation}\label{coef} 
 y^{2k}~ {{\zeta (2k + 2s + 1, {1 \over 2})} \over {\Gamma(s)}}~,
 ~ k = 0,1,2,... ~~.
\end{equation}
Then by adjusting the value of $s$ to $s=-1$ the term proportional 
to $y^2$ indeed leads to a finite nonvanishing value for $m^2\to 0$.
In addition all the other coefficients for $k\neq 1$ vanish
due to the pole of $\Gamma (s)$ at $s=-1$.
From Eqs.~(\ref{zetaexpand}) and (\ref{zetas0})
we calculate 
\begin{equation}\label{zetas-1} 
  \zeta_{s = -1} (x) = - \zeta(3,{1 \over 2})~y^2~\zeta_{s = 0} (x)~,
\end{equation} 
using the definitions given in Eqs.~(\ref{Iexpand},\ref{N}).

The effective lagrangian for $\pi^0\sigma \to 2\gamma$ thus becomes 
\begin{equation}\label{L2ga2}
 {\cal L}_{T}^{2 \gamma}
 \simeq  {{N_C~ e^2} \over {16 \pi^2}} \zeta (3, {1 \over 2})~
\epsilon^{\mu \nu \alpha \beta}~F_{\mu \nu}~F_{\alpha \beta}~ 
   {\lim_{m \to 0} } 
         ~({m \over {2 \pi T}})^2 
Tr \left( \frac{\pi^0\tau_3 Q^2}{2[m/g+\sigma]} \right) ~~.
\end{equation}
Keeping the term linear in the $\sigma$ field the $m\to 0$ limit is defined,
\begin{equation}\label{L2ga3}
 {\cal L}_{T}^{2 \gamma}
 =  -  {{ 7 \zeta (3)~ g^2 \alpha N_C} \over {96 \pi^3~T^2}}~ 
\epsilon^{\mu \nu \alpha \beta}~F_{\mu \nu}~F_{\alpha \beta}~
 ( \pi^0~\sigma )~.
\end{equation}
We note that the effective photon coupling to neutral particles
$(\pi^0,\sigma)$ given in Eq. (\ref{L2ga3}) is manifestly gauge invariant.
By comparing the above coupling at high $T$ with the one
at $T = 0$  \cite{rb,rg} we note the substitution 
\begin{equation}\label{FpiT}
 { 1 \over {F_\pi}} \rightarrow 
  { {7 \zeta (3) g^2} \over {4 \pi^2 T^2} }~\sigma~,
\end{equation}
which is in agreement with the derivation given in \cite{pisa1}
based on evaluating triangle and box diagrams, respectively.
Actually it is this relation which allows to fix the
normalization factor $N (s)$ of Eq.~(\ref{N}) by the
requirement: 
\begin{equation}\label{Nfix} 
T~ \Sigma \int~ {{d^3p} \over {(2\pi)^3}} 
{ 1 \over {(p_E^2 + m^2)^{2 - s}}}~~
  {\buildrel {T \gg m} \over {=} } ~ N (s)~ ( {T \over m })^{2s}~
 \int~ {{d^4p} \over {(2\pi)^4}} 
{ 1 \over {(p_E^2 + m^2)^{2 - s}}}~.
\end{equation}

In a corresponding fashion the anomalous couplings 
linear in $F_{\mu \nu}$ for e.g. $\gamma ~\to 3 \pi~ \sigma$
may be derived, which turn out to be in leading order
(keeping $s = - 2$)
proportional to $\zeta (5)~g^4/T^4$ at high temperature.

Generalizing the result of Eq.~(\ref{L2ga3})
to the case of $SU (3)$ 
($\vec\tau$ matrices are replaced by the Gell-Mann matrices 
$\vec\lambda$, the pion field $\vec\pi$ by the eight pseudoscalar bosons
$\vec\phi = \pi, K, \eta_8$, correspondingly)
allows to obtain the 
radiative two-photon couplings for 
the processes $\eta~(\eta^{\prime})~\sigma \rightarrow 2 \gamma$
in the chirally symmetric limit
(besides the corresponding $T = 0$ normalization and the
$\eta - \eta^{\prime}$ mixing \cite{donoh}).
There are anomalous couplings between photons
and mesons with more than one $\sigma$ field, which, however, we do 
not consider here.

%%%%%%%%%%%%%%%%%%%%%%%%%%%%%%%%%%%%%%%%%%%%%%%%%%%%%%%%%%%%%

\section{Anomalous hadronic couplings} \label{had}

We now consider the anomalous couplings of interacting bosons at
high temperature in the chirally symmetric phase.
Following the expansion of the heat kernel, Eq.~(\ref{zetaexpand}),
these interactions are contained in the terms of $O(d^4)$, namely:
$ (d \cdot d)^2, ~ d_{\mu} (d \cdot d) d_{\mu}$ and
$ d_{\mu} d_{\nu} d_{\mu} d_{\nu}$, respectively,
where $d_{\mu} \sim \sigma_{\mu \nu} {\overline A}_{\nu} \gamma_5$
with ${\overline A}_{\mu}$ given by the $SU(3)$ version of Eq.~(\ref{A}).

It is useful to summarize shortly the $T=0$ case:
in the broken phase the simplest well known example of hadronic
interactions is the one for
$K^+ ~ K^- \rightarrow \pi^+ ~\pi^- ~\pi^0$. 
After performing the Dirac trace (cf. Eq.~(\ref{tr})),
and the Taylor expansion in the variable $t$, the leading
term has the following structure\cite{donoh} 
\begin{equation}\label{WZWT0} 
 {\cal L}_{T=0} \simeq {{8 N_C} \over {12 \pi^2}}
\int_0^1 ~dt ~t^4~ Tr ~\bigl( 
\overline{\varphi}~ \epsilon^{\mu \nu \rho \sigma}
 ~ {\tilde A}_{\mu} {\tilde A}_{\nu} 
{\tilde A}_{\rho} {\tilde A}_{\sigma} \bigr)~,
\end{equation}
where $Tr$ denotes the $SU(3)$ flavour trace,
with $\overline{\pi} \to \overline{\varphi} \equiv \vec{\lambda} \cdot
\vec{\varphi}/ 2 F_{\pi}$ and ${\tilde A}^{\mu} \equiv \partial^{\mu}
\overline{\varphi}$ for $SU(3)$.

In order to describe $K K \to \pi \pi \pi \sigma$ interactions
at $T \ne 0$ \cite{pisa1,pisa2} in the symmetric phase we have to find
the nonvanishing - finite -  terms of $\Gamma$, Eq.~(\ref{WZW}), i.e. of 
$\zeta_s (x)$ in Eq.~(\ref{zetaexpand}) in the limit $m \to 0$.
Here the lowest power in $1/m^2$ from the product of $\bar\varphi$
and the fourth power in the axial current ${\overline A}_\mu$ of Eq.~(\ref{A})
present in the $O(d^4)$ terms is $1/m^6$. So by adjusting the value of $s$
to $s=-3$ the term proportional to $y^6$ in  (\ref{zetaexpand})
leads to a finite nonvanishing limit for $m\to 0$. Again all the 
other contributions vanish due to the pole of $\Gamma (s)$ at 
$s=-3$ (cf. Eq.~(\ref{coef})).
We calculate   
\begin{equation}\label{zetas-3}
 \zeta_{s=-3}(x) =  \frac{15}{8}\zeta(7,\frac{1}{2})~ y^6 ~
 \zeta_{s=0}(x) ~.
\end{equation}
Expanding in $t$ as for $T=0$ ~$\zeta_{s=0}(x)$ is expressed in terms
of the current ${\overline A}_\mu$, Eq.~(\ref{A}), and
we find the leading hadronic interaction linear in the $\sigma$ field
at high $T$ given by 
%%%%
\begin{eqnarray}\label{WZWT}
 {\cal L}_{T} & \simeq & \frac{127~\zeta (7)}{128}\frac{N_C}{\pi^2} 
 ~({g \over {2 \pi T}})^6
\nonumber \\
 & \times & Tr~~ (\vec{\lambda} \cdot \vec{\varphi})~
    \epsilon^{\mu \nu \rho \tau}~
  \left\{~~ \left[ \left(  
  \sigma  \partial_{\mu} (\vec{\lambda}  \cdot \vec{\varphi})
           - (\vec{\lambda} \cdot \vec{\varphi}) \partial_{\mu} \sigma 
 \right)
~ \partial_{\nu} (\vec{\lambda} \cdot \vec{\varphi})~
 \partial_{\rho} (\vec{\lambda} \cdot \vec{\varphi})~
 \partial_{\tau} (\vec{\lambda} \cdot \vec{\varphi})~
 \right. \right. \nonumber \\
 & & \left. + \mbox{\quad cyclic permutations of\quad} (\mu,\nu,\rho,\tau)
 \right] \nonumber\\
 & & \left. - 9 \sigma \partial_{\mu} (\vec\lambda\cdot\vec\varphi )
  \partial_{\nu}(\vec\lambda\cdot\vec\varphi ) 
  \partial_{\rho} (\vec\lambda\cdot\vec\varphi ) 
  \partial_{\tau}(\vec\lambda\cdot\vec\varphi ) 
 \right\}~~ .
\end{eqnarray}
The value of the zeta-function is $\zeta(7) = 1.00835$ \cite{abram}.
This is the effective lagrangian at high $T$ in the symmetric phase,
which replaces the one of Eq.~(\ref{WZWT0}) at $T=0$.
${\cal L}_{T}$ shows a strong suppression with
increasing temperature (for fixed coupling $g$).

Other anomalous interactions with more than one $\sigma$ field
may be worked out in a similar way; but again this requires
the proper adjustment of the parameter $s$ when performing the
symmetric limit $m^2 \to 0$.
At high $T$ we do not find an universal ${\cal L}_{odd}$
which can be given in a simple compact form, comparable to
the Wess-Zumino-Witten action $\Gamma_{WZW}$ at $T=0$ \cite{rg}.

In the spirit of effective lagrangians derived in the hard thermal loop 
expansion \cite{HTL} Pisarski argues in \cite{pisa2} that 
local anomalous interactions as given by Eq. (\ref{WZWT})
are strictly valid only in the static limit \cite{static}.
The nonstatic case is expected to be more complicated than Eq. (\ref{WZWT})
because nonlocal contributions have to be included at high $T$.
  
%%%%%%%%%%%%%%%%%%%%%%%%%%%%%%%%%%

\section{Conclusions} \label{conc}

Finally, one may ask about phenomenological consequences,
i.e. possible  observable effects, of these high temperature
anomalous couplings.
An interesting case for testing Eq.~(\ref{L2ga3}) -
and those related to the anomalous interactions of vector
mesons at high $T$ \cite{pisa1,pisa2} - 
are photon- and dilepton measurements.

In this context we mention the production of
low-mass lepton pairs in heavy ion collisions at
 high energies\cite{ceres,helios}.
In order to interpret these data, especially for masses 
below and near the $\rho$ meson mass,
a detailed knowledge of the electromagnetic interactions
of $\pi^0, \eta, \eta^{\prime}, \omega, \rho,$
etc. in a high density hadronic environment (i.e. maybe
in a chirally symmetric phase at high $T$)
is important.

As a preliminary  example based on the coupling of Eq.~(\ref{L2ga3})
we  estimate the production rate of
dileptons from the thermal production due to the process 
$\sigma \pi \rightarrow \gamma^{\star} \gamma
\rightarrow e^+e^-~\gamma$ by using a few simplifying
assumptions like e.g. vanishing $\pi$ and $\sigma$ masses, 
thermal equilibrium with Boltzmann distributions.
In order to quote a temperature independent quantity
we relate the resulting rate to the one calculated
from the $\rho$ dominated thermal process: $\pi~ \pi \rightarrow
e^+~e^-$ by evaluating both rates at the $\rho$ peak,
i.e. for $m_{e^+e^-} \simeq m_{\rho}$.
The ratio is  extremely small, it is of the order:
$10^{-4}~\alpha ~g^4~(\Gamma_\rho / m_\rho)^2)$.

%%%%%%%%%%%%%%%%%%%%%%%%%%%%%%%%%%%%%%%%%%%%%%%%%%%%%%%%%%%%%
%\section{Summary}
%%%%%%%%%%%%%%%%%%%%%%%%%%%%%%
%\newpage  
%%%%%%%%%%%%%%%%%%%%%%%%%%%%%%%%%%%%%%%%%%%%%%%%%%%%%%%%%%%%%%%%%%%%%
\section*{Acknowledgements}

We like to thank R. D. Pisarski for comments and I. Dadi\'c for 
discussions.
%%%%%%%%%%%%%%%%%%%%%%%%%%%%%%%%%%%%%%
Partial support by the EEC Programme "Human Capital
and Mobility", Network "Physics at High Energy Colliders",
Contract CHRX-CT93-0357
is acknowledged .
%%%%%%%%%%%%%%%%%%%%%%%%%%%%%%%%%%%%%%%%%%%%%%%%%%%%%%%%%%%%%%%%%%%%%%
\newpage
%%%%%%%%%%%%%%%%%%%%%%%%%%%%%%%%%%%%%%%%%%
%\Section{References}

%%%%%%%%%%%

\begin{references}
\bibitem{static}
After taking the different normalization factors of the $\sigma$ field
($\sigma$ vs. $\sqrt{2/3}~\sigma$ ) and the SU(3) matrices $\vec\lambda$
into account there still remains a factor 5 discrepancy in the 
langrangian density for one $\sigma$ meson with zero momentum,
i. e. for $\partial_\mu\sigma =0$, following from Eq. (\ref{WZWT})
and the one given in \cite{pisa2}.
Choosing e.g. $\mu=0$ we obtain
\begin{displaymath}
 -5~ \frac{127\zeta (7) g^6 N_C}{256 \pi^8 T^6} ~
 \sigma ~\epsilon^{ijk}~ Tr~ ( \frac{\vec\lambda}{2}\cdot \vec\varphi)~
  \partial_0 ( \frac{\vec\lambda}{2}\cdot \vec\varphi )~
  \partial_i ( \frac{\vec\lambda}{2}\cdot \vec\varphi )~
  \partial_j ( \frac{\vec\lambda}{2}\cdot \vec\varphi )~
 \partial_k ( \frac{\vec\lambda}{2}\cdot \vec\varphi ) ~~.
\end{displaymath}
\end{references}
\end{document}